\documentclass[a4paper]{spie}  
\usepackage[]{graphicx}            
\usepackage{multirow}

\def\CN2{\mbox{$C_N^2 \ $}}

\def\CT2{\mbox{$C_T^2 \ $}}
\def\tauO{\mbox{$\tau_{0} \ $}}
\def\thetaO{\mbox{$\theta_{0} \ $}}

\def\sigmal2{\mbox{$\sigma ^{2}_{I} \ $}}

\title{MOSE: a feasibility study for optical turbulence forecasts with the Meso-Nh mesoscale model to support AO facilities at ESO sites (Paranal and Armazones).} 

\author{Elena Masciadri\supit{a}, Franck Lascaux\supit{a}
\skiplinehalf
\supit{a}INAF - Osservatorio Astrofisico di Arcetri, L.go E. Fermi 5, 50125  Florence, Italy\\
}

\authorinfo{Correspondence to Elena Masciadri: masciadri@arcetri.astro.it}

\begin{document} 
  \maketitle 

\begin{abstract}
We present very encouraging preliminary results obtained in the context of the MOSE project, an on-going study aiming at investigating the feasibility of the forecast of the optical turbulence and meteorological parameters (in the free atmosphere as well as in the boundary and surface layer) at Cerro Paranal (site of the Very Large Telescope - VLT) and Cerro Armazones (site of the European Extremely Large Telescope - E-ELT), both in Chile. The study employs the Meso-Nh atmospheric mesoscale model and aims at supplying a tool for optical turbulence forecasts to support the scheduling of the scientific programs and the use of AO facilities at the VLT and the E-ELT. In this study we take advantage of the huge amount of measurements performed so far at Paranal and Armazones by ESO and the TMT consortium in the context of the site selection for the E-ELT and the TMT to constraint/validate the model. A detailed analysis of the model performances in reproducing the atmospheric parameters (T, V, p, H, ...) near the ground as well as in the free atmosphere, is critical and fundamental because the optical turbulence depends on most of these parameters. This approach permits us to provide an exhaustive and complete analysis of the model performances and to better define the model operational application. This also helps us to identify the sources of discrepancies with optical turbulence measurements (when they appear) and to discriminate between different origins of the problem: model parameterization, initial conditions, ... Preliminary results indicate a great accuracy of the model in reproducing most of the main meteorological parameters in statistical terms as well as in each individual night in the free atmosphere and in proximity of the surface. The study is co-funded by ESO and INAF-Arcetri (Italy). 
\end{abstract}



\section{INTRODUCTION}
\label{sec:intro} 

The MOSE project (MOdeling ESO Sites) aims at proving the feasibility of the forecast of the optical turbulence OT ($\CN2$ profiles) and all the main integrated astro-climatic parameters derived from the $\CN2$ i.e. the seeing ($\varepsilon$), the isoplanatic angle ($\thetaO$), the wavefront coherence time ($\tauO$) above the two ESO sites of Cerro Paranal (site of the Very Large Telescope - VLT) and Cerro Armazones (site selected for the European Extremely Large Telescope - E-ELT). The OT forecast is a crucial cornerstone for the feasibility of the ELTs: it is fundamental for supporting all kind of AO facilities in an astronomical observatory and for performing the flexible-scheduling of scientific programs and instrumentation through the Service Mode. The MOSE project aims at overcoming two major limitations that we normally encounter in studies focused on the optical turbulence forecast with atmospheric models: {\bf (1)} the difficulty in having independent samples of measurements for the model calibration and model validation to estimate if and how the correlation between measurements and predictions decreases with the increasing of the number of nights used for the calibration; {\bf (2)} the difficulty in having a large number of simultaneous measurements done with different and independent instruments for the OT estimates (in particular vertical profilers). This project is performed with the non-hydrostatic mesoscale atmospherical models Meso-Nh\cite{lafore98} joined with the Astro-Meso-Nh package for the calculation of the optical turbulence\cite{masciadri99a,masciadri99b} to perform the OT forecasts. An extended data-set of observations (meteorological parameters and optical turbulence) have been considered in the project. In this contribution we focus our attention on the model performances in reconstructing the meteorological parameters near the surface as well as in the free atmosphere. 

\section{WHOLE OBSERVATIONS DATA-SET}
\label{sec:obs} 
At Paranal, observations of meteorological parameters near the surface come from an automated weather station (AWS) and a 30~m high mast including a number of sensors at different heights. Both instruments are part of the VLT Astronomical Site Monitor \cite{vlt99}. Absolute temperature data are available at 2~m and 30~m above the ground.
Wind speed data are available at 10~m and 30~m above the ground. At Armazones, observations of the meteorological parameters near the ground surface come from the the Site Testing Database \cite{Schoeck2009},  more precisely from an AWS and a 30~m tower (with temperature sensors and sonic anemometers). Data on temperature and wind speed are available at 2~m, 11~m, 20~m and 28~m above the ground. At 2~m (Armazones) temperature measurements from the AWS and the sonic anemometers are both available but we considered only those from the tower (accuracy of 0.1$^{\circ}$C)\cite{Skidmore2007}. Those from the AWS are not reliable because of some drift effects (T. Travouillon, private communication). Wind speed observations are taken from the AWS (at 2~m) and from the sonic anemometers of the tower (at 11~m, 20~m and 28~m). The outputs are sampled with a temporal frequency of 1 minute. 

At Paranal we access also to 50 radio-soundings (vertical distribution of the meteorological parameters in the $\sim$ 20~km above the ground) launched above this site in the context of an intense site testing campaign for water vapor estimates\cite{chacon2011} and covering 23 nights in 2009, 11 nights in summer and 12 in winter time. In a subsample of these nights (16), a few radio-soundings (two or three) have been launched at different times in the same night. 

Observations of the optical turbulence at Paranal, relate to the Site Testing Campaign of November-December 2007\cite{Dali2010}, come from a Generalized Scidar, a DIMM and a MASS. The Generalized Scidar measurements have been recently re-calibrated\cite{masciadri2012}. Optical turbulence measurements at Armazones come from a DIMM and a MASS\cite{Schoeck2009} that have been used for the TMT site selection campaign.

\section{MODEL CONFIGURATION}
\label{sec:mod_conf} 
\begin{figure}
\centering
\includegraphics[width=\textwidth]{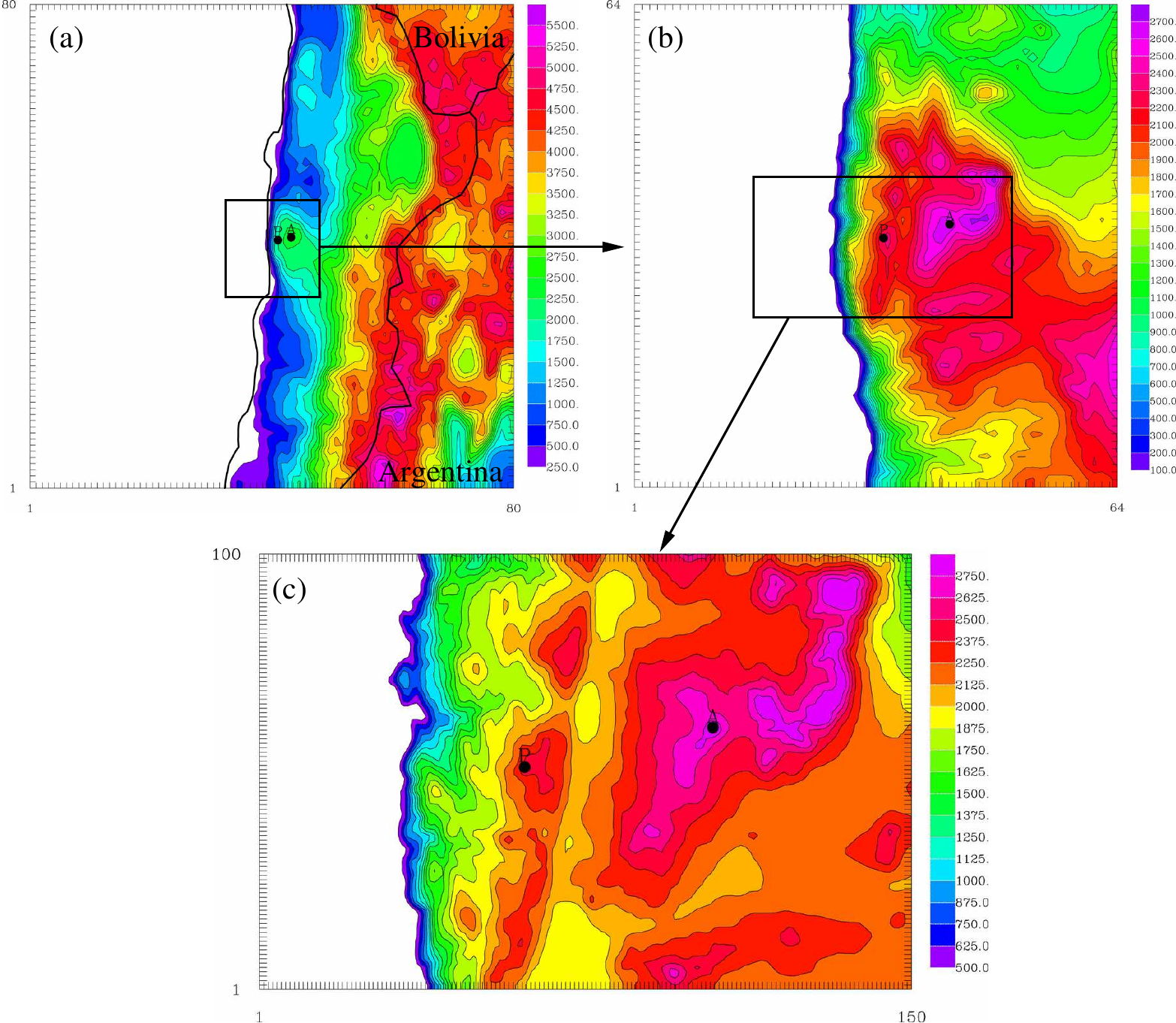}
\caption{\label{pgd} Orography of the region of interest as seen by the Meso-Nh model (polar stereographic projection) for all the imbricated domains of the grid-nesting configuration: (a) Domain 1, (b) Domain 2, (c) Domain 3. The black dots report the position of Cerro Paranal and Cerro Armazones. See Table \ref{tab_orog} for specifications of the domains (number of grid-points, domain extension, horizontal resolution). }
\end{figure}

The Meso-Nh atmospherical mesoscale model can simulate the temporal evolution of three-dimensional meteorological parameters over a selected finite area of the globe.
The system of hydrodynamic equations is based upon an anelastic formulation allowing for an effective filtering of acoustic waves.
It uses the Gal-Chen and Sommerville\cite{Gal75} coordinates system on the vertical and the C-grid in the formulation of 
Arakawa and Messinger\cite{Arakawa76} for the spatial digitalization.
It employs an explicit three-time-level leap-frog temporal scheme with a time filter \cite{Asselin72}.
For this study we use a 1D mixing length proposed by Bougeault and Lacarr\`ere \cite{Bougeault89} with a one-dimensional 1.5 turbulence closure scheme \cite{Cuxart00}. The surface exchanges are computed by the Interaction Soil Biosphere Atmosphere - ISBA scheme\cite{Noilhan89}.
The grid-nesting technique \cite{Stein00}, employed in our study, consists of using different imbricated domains of the Digital Elevation Models (DEM i.e orography) extended on smaller and smaller surfaces, with increasing horizontal 
resolution but with the same vertical grid. The standard configuration of this study includes three domains (Fig.\ref{pgd} and Table \ref{tab_orog}) with the lowest horizontal resolution equal to 10~km and the highest horizontal resolution equal to 0.5~km. The orographic DEMs that we used for this project are the GTOPO\footnote{$http://www1.gsi.go.jp/geowww/globalmap-gsi/gtopo30/gtopo30.html$} with an intrinsic horizontal resolutions of 1~km (used for the domains 1 and 2) and the ISTAR\footnote{Bought by ESO at the ISTAR Company - Nice-Sophia Antipolis, France} with an intrinsic horizontal resolution of 0.5~km (used for the domain 3). Along the z-axis we have 62 levels distributed as the following: the first vertical grid point equal to 5~m, a logarithmic stretching of 20~$\%$ up to 3.5~km above the ground, and an almost constant vertical grid size of $\sim$600~m up to 23.8~km. The model has been parallelized using OPEN-MPI-1.4.3 and it run on local workstations and on the HPCF cluster of the European Centre for Medium weather Forecasts (ECMWF). The second solution permitted us to achieve relatively rich statistical estimates of these analysis.

\begin{table*}
\begin{center}
\caption{\label{tab_orog} Orography: grid-nesting configuration conceived as three imbricated domains (1, 2 and 3) with a horizontal resolution from a minimum of 10~km to a maximum of 0.5~km (column 2) extended on smaller and smaller domains (column 4) with a different number of grid points (column 3). See Fig.\ref{pgd}.}
\vskip 0.2cm
\begin{tabular}{|c|c|c|c|}
\hline
Domain & $\Delta$X & Grid Points & Surface \\
& (km) & & (km$\times$km) \\
\hline
Domain 1       & 10& 80$\times$80& 800$\times$800\\
Domain 2 & 2.5 &  64$\times$64&   160$\times$160 \\
Domain 3  & 0.5 &  150$\times$100& 75$\times$50  \\
\hline
\end{tabular}
\end{center}
\end{table*}

\section{COMPARISON MESO-NH VERSUS OBSERVATIONS}
\label{sec:res}

To estimate the statistical model reliability in reconstructing the main meteorological parameters we used the averaged values plus two statistical operators: the bias and the root mean square error (RMSE) defined as:

\begin{equation}
BIAS = \sum\limits_{i = 1}^N {\frac{{(Y_i  - X_i )^{} }}
{N}} 
\label{eq1}
\end{equation}

\begin{equation}RMSE = \sqrt {\sum\limits_{i = 1}^N {\frac{{(Y_i  - X_i )^2 }}
{N}} } 
\label{eq2}
\end{equation}
where $X_{i}$ are the individual observations,  $Y_{i}$ the individual simulations calculated at the same time and N is the number of times in which a couple ($X_{i}$, $Y_{i}$) is available and different from zero for each time. At the same time, due to the fact that we are interested in investigating the model ability in forecasting a parameter and not only in characterizing it, it is important for us to investigate also the correlation observations/simulations calculated night by night and not only in statistical terms. This further detailed analysis has been performed for the potential temperature and the wind speed only i.e. the two most important parameters from which the optical turbulence depends on. In relation to this last issue, in this contribution we present only the results obtained with the temperature because of lacking of space.

\subsection{Vertical distribution of the meteorological parameters}
\label{sec:vert_dist} 
\begin{figure}
\centering
\includegraphics[width=14cm]{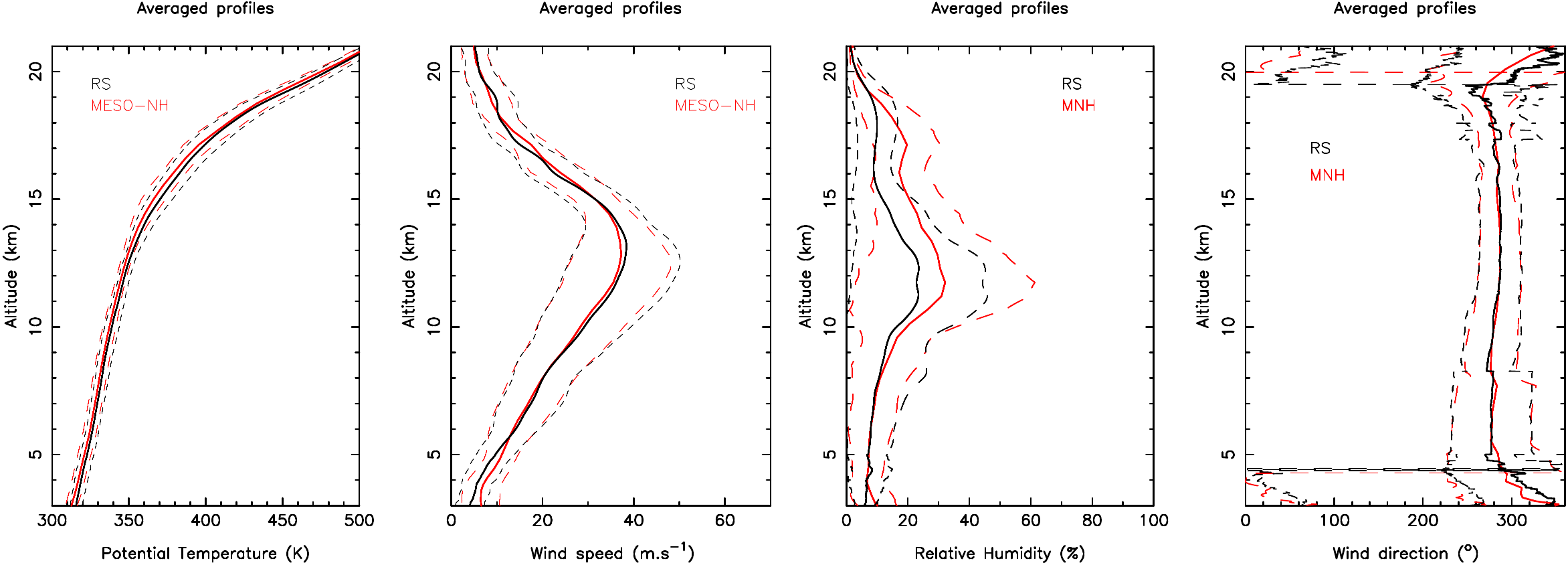}
\caption{\label{average} Averaged profiles observed by radio-soundings and simulated by the Meso-Nh mesoscale model on the sample of {\bf 50 flights}. The dashed line indicates the standard deviation. The wind direction is reported as modulus 360$^{\circ}$ therefore the large $\sigma$ in proximity of the ground (at $\sim$4~km) and in the very high part of the atmosphere (at $\sim$19~km) is just fictitious. } 
\end{figure}

\begin{figure}
\centering
\includegraphics[width=14cm]{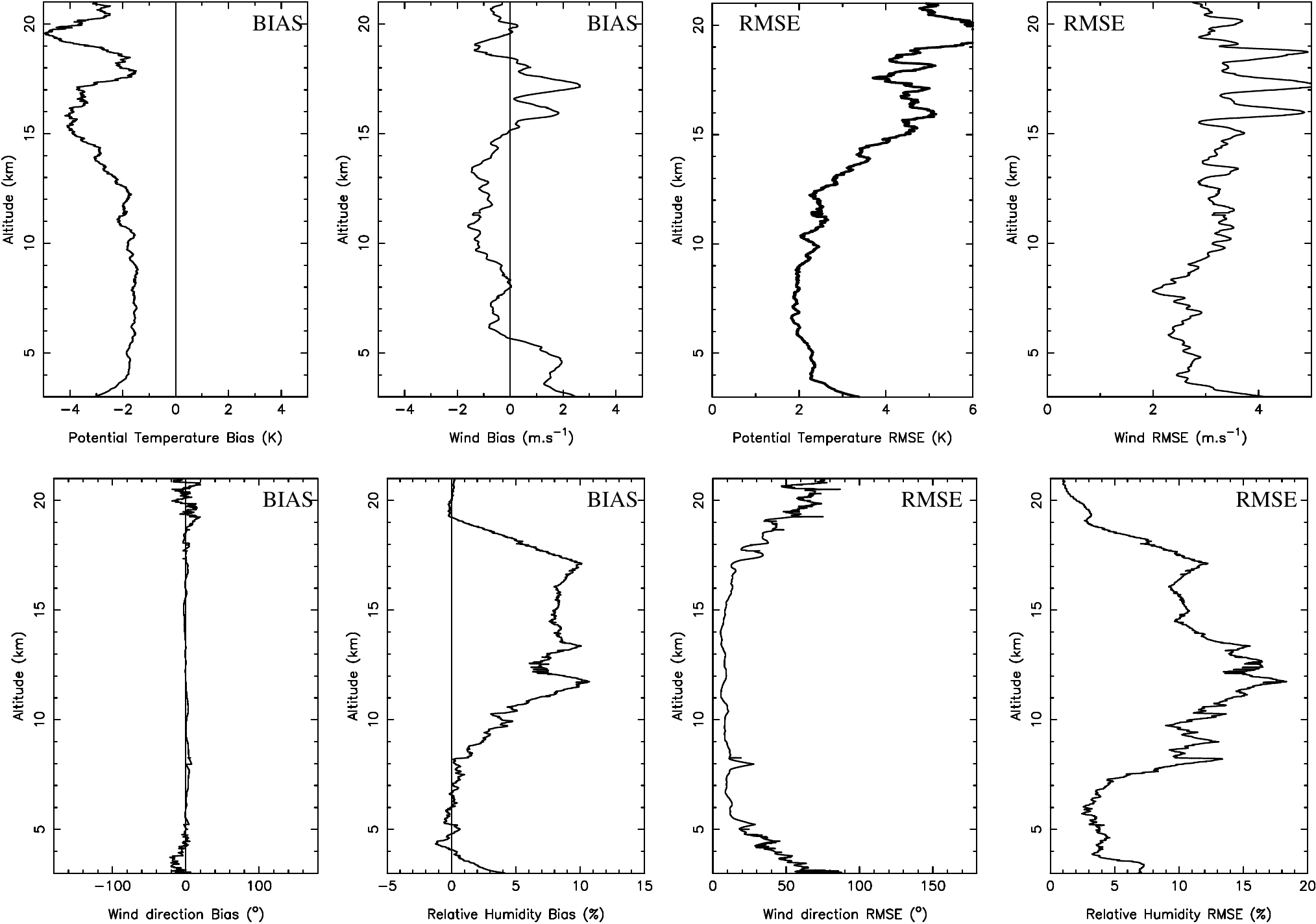}
\caption{\label{wind_and_co} {\bf BIAS} (simulations minus observations) and {\bf RMSE} calculated for the vertical profiles of potential temperature, wind speed, wind direction and relative humidity related to a sample of {\bf 50 flights}.} 
\end{figure}

To quantify the model statistic reliability in reconstructing the vertical distribution of the meteorological parameters we compare observations from the radio-soundings ({\bf 50 flights}) with the simulations performed by the model in the range [3~km - 21~km]. The first 400-500 meters above the ground (h=2634~m - Paranal summit) constitute, indeed, a sort of 'gray zone' in which it can be meaningless to retrieve any quantitative useful estimates from the radio-soundings because of many different reasons. Among others: (a) the orographic maps of the innermost domain has an intrinsic $\Delta$h$\sim$156~m with respect to the real summit due to the natural smoothing effect of the model horizontal interpolation of the DEM, (b) the radio-soundings have been launched at around 50~m below the summit. It should be meaningless to compare the observed and simulated values at the summit ground height because in one case we resolve friction of the atmospheric flow near the ground, in the other no, (c) we have to take care about an uncertainty $\Delta$h of around 50~m in the identification of the zero point (h$_{0}$) probably due to an unlucky procedure performed during the radio-soundings launches on the the zero point setting. This uncertainty has basically no effects above a few hundreds of meters above the ground because the parameters values are affected by phenomena evolving at larger spatial scales.  We decided therefore to treat data only above roughly 500~m from the summit. Two different treatments of the model outputs have been analyzed: {\bf (1)} we take the vertical profile calculated by the model at the exact instant of the launch time; {\bf (2)} we take the averaged vertical profiles simulated by the model in around one hour from the time in which the radio-sounding has been launched and one hour later. We considered that the balloon is an in-situ measurement and a balloon needs around 1 hour to cover 20~km from the ground moving up in the atmosphere with a typical vertical velocity of $\sim$6~m$\cdot$s$^{-1}$. The balloon have been launched close to the synoptic hours (00:00 UT, 06:00 UT, 12:00 UT).
The model outputs as well as the balloons observations are re-interpolated with a constant vertical grid with size equal to the first vertical grid point (5~m) of the model before to be compared. Fig.\ref{average} reports the averaged values calculated on the sample of 50 flights. Fig.\ref{wind_and_co} show the results of bias and RMSE calculated on the same sample of flights. Results obtained with approach (1) and (2) are substantially the same therefore we report just the results of one case. The bias contains informations on systematic model off-sets. The RMSE tells us the maximum dispersion achieved by the model on the whole sample.  

Looking at Fig.\ref{average} and Fig.\ref{wind_and_co} we retrieve that the Meso-Nh model shows very good performances in reconstructing the wind speed direction on the whole 20~km. Between 5~km and 18~km the model reconstruction is statistically almost perfect. Below 5~km (where the orographic effects are most evident) the bias is of the order of $\sim$20$^{\circ}$ with a RMSE that can reach a few tens of degrees. This means that, night by night, in the very proximity of the surface we can have a discrepancy of the order of a few tens of degrees. It is worth to highlight that the accuracy in observing the wind direction can be hardly be better than $\sim$20$^{\circ}$. The model performances are therefore very satisfactory. The model shows very good performances in reconstructing the relative humidity. It is basically never larger than 10$\%$ all along the 20~km. The largest discrepancy (10$\%$) of simulations with respect to measurements is observed at the jet stream level. Such a satisfactory result has been obtained in spite of the fact we used a cheap scheme (in terms of CPU cost) for the relative humidity. That was possible because of the dryness of the region. Such a solution permits faster simulations. The small bump at the jet-stream level (Fig.\ref{average}) is highly probably due to the humidity coming from the close ocean. The model shows also a very good performances in reconstructing the potential temperature. We observe a very small bias of $\sim$ 2$^{\circ}$C from the ground up to around 13~km. Above 13~km, where the potential temperature slope is steeper and steeper, the bias can achieve up to 4$^{\circ}$C. The wind speed intensity is very well reconstructed: we have a bias of around 1~m$\cdot$s$^{-1}$ in the [5~km - 15~km] range for the wind speed. Above 15~km and in the [3~km - 5~km] range the bias achieve a value of 2~m$\cdot$s$^{-1}$. Comparing results (not shown here) obtained with the Meso-Nh model and the ECMWF analyses coming from the General Circulation Models (CGM) we could conclude that most of the residual biases and RMSEs we described so far are generated by initial conditions and not by the mesoscale model itself. 

\begin{figure}
\centering
\includegraphics[width=14cm]{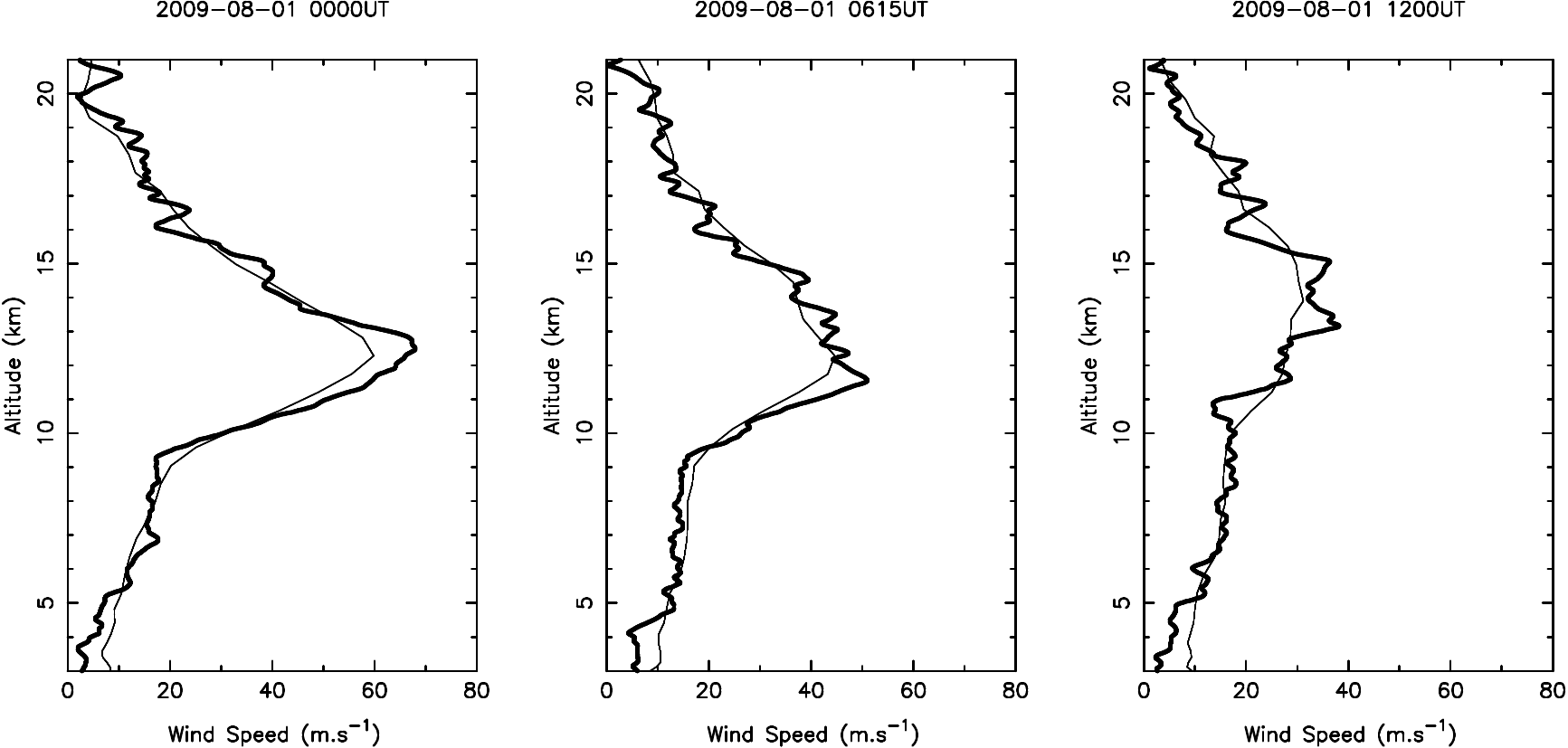}
\caption{\label{radio_mnh} Comparison of wind speed intensity observed (radio-soundings: thick line) and simulated (Meso-Nh model: thin line) at three different instants (00:00, 06:15, 12:00 UT) during the same night: 1/8/2009 above Paranal.} 
\end{figure}

A comparison (observations/simulations) performed night by night and in each instant for which a radio-soundings was available revealed an excellent agreement in basically all the 50 cases studied. This analysis has been performed for the wind speed and the potential temperature, the two main parameters from which the optical turbulence depends on. As an example, in Fig.\ref{radio_mnh} is shown a comparison of the wind speed observed and simulated in three different instants (00:00, 06:15 and 12:00 UT) of the same night. This case perfectly puts in evidence the excellent performances of the model in adapting itself to the wind speed evolution during the time. In spite of the fact that the observed wind speed strongly modifies its features all along the night at different heights, we note that the model perfectly reconstructs the observed wind speed features in the three different instants. A similar behavior is observed in basically all the 50 cases studied (details in a forthcoming report of the MOSE Project). 

This definitely guarantees us the reliability of a tool (the Meso-Nh mesoscale model) to reconstruct the temporal evolution of the vertical distribution of the wind speed (V(h,t)) during a whole night. This is a fundamental ingredient (beside to the vertical profiles of the optical turbulence $\CN2$(h,t)) to be used for the calculation of the temporal evolution of the wavefront coherence time $\tauO$(t):
\begin{equation}
\tau _0 (t) = 0.057 \cdot \lambda ^{6/5} (\int\limits_0^\infty  {V(h,t)^{5/3} }  \cdot C_N^2 (h,t)dh)^{ - 3/5} 
\end{equation}

We intend here to stress the concept that the analyses and the forecasts of the meteorological parameters retrieved from the General Circulation Models (ECMWF or NOAO) are calculated only at synoptic hours (00:00, 06:00, 12:00 and 18:00) UT. 
\begin{figure}
\centering
\includegraphics[width=14cm]{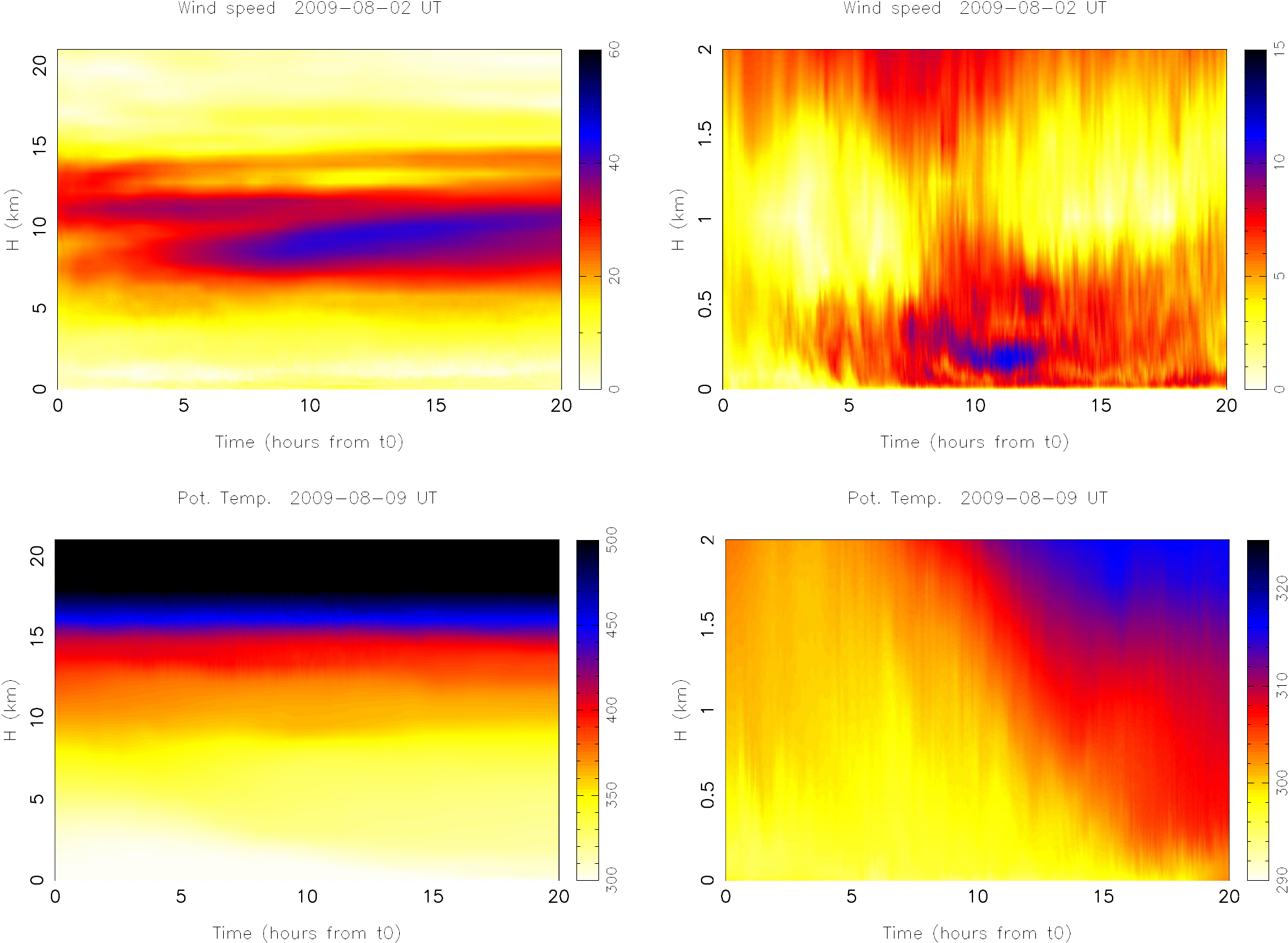}
\caption{\label{temp_evol} Temporal evolution of the wind speed (top) and potential temperature (bottom) vertical distribution, calculated on the grid point of Paranal and extended 
along 21~km (left) and 2~km (right) from the ground. The simulation starts at t$_{0}$=18 UT and lasts 20 hours. The local night (20:00 - 05:00 LT) corresponds to the interval (6 - 15) on the x-axis.} 
\end{figure}
Fig.\ref{temp_evol} shows the temporal evolution of the wind speed and the potential temperature provided by the Meso-Nh model during two different nights. In this example we can appreciate the intrinsic level of the temporal variability of both parameters at different heights above the ground during the night. This is far from being negligible. In other words, the mesoscale predictions provide us a complete information (temporal evolution of the metorological parameter) with respect to the estimates coming from the General Circulation Models that can provide outputs only at synoptic hours and can not fill the lacking information in between the synoptic hours. This proves us the invaluable utility of a mesoscale model for the prediction of the astro-climatic parameters. In particular, our results indicate that the wind speed retrieved from the mesoscale model is, at present, certainly the best (and the most practical) solution to calculate the temporal evolution of the wavefront coherence time $\tauO$(t). The temporal evolutions of the potential temperature and the wind speed contribute both in determining the prediction of the optical turbulence $\CN2$(h,t).

A few more words are suitable to comment the wind speed in the vertical slab [3~km - 5~km] a.s.l. that is [0.5~km - 2.5~km] a.g.l.. The discrepancy observation/simulation of around 1-2~m$\cdot$s$^{-1}$ is weak but, differently from the other discrepancies, seems to be the only one derived directly by the mesoscale model and not by the initial conditions. Even if the discrepancy is weak, it is therefore worth to analyze more in detail that. We counted a roughly comparable number of cases in which the model overestimates and correctly estimates the wind in this vertical slab. The model almost never underestimates the observed values. We did not observe any clear correlation between the discrepancy with the absolute value of the wind speed. We tested the sensitivity to the grid-point selection to check if a not precise selection of the grid-point of the summit could create some anomalous effects on the wind in the low atmosphere: for all the cases in which an overestimate has been observed we calculated the same bias in four different grid-points around the summit but it has not been observed any substantial difference. We note that the radio-sounding is an in-situ measurement and that the balloon moves horizontally along the (x,y) plan during the ascension in the atmosphere. During the ascension time, it therefore senses a volume of atmosphere shifted with respect to the zenithal direction. The radio-sounding lasts around 4 minutes with a V$_{z}$=6~m$\cdot$s$^{-1}$ to achieve the altitude of 4~km a.s.l. In this temporal interval the balloon can move somewhere (depending on the wind direction) within circle with a radius of ~2.4~km. If we calculate the maximum variation of the wind speed ($\Delta$V) inside such a circle we see that, at 4~km a.s.l., $\Delta$V is of the order of 1.5-2 m$\cdot$s$^{-1}$. Table \ref{3_5_wind_speed} reports these values for a few flights. The inhomogeneity decreases with the height and disappears in the high part of the atmosphere. In other words, being that the horizontal distribution on the (x,y) plane of the wind speed in the low part of the atmosphere is not necessarily homogeneous, this could explain the discrepancy with the simulations in the [3~km-5~km] range. This argument tells us that the radio-sounding is not an optimal reference for comparisons with simulations to be done in the low part of the atmosphere.  A preferable choice should be an instrument based on an optical remote sensing principle. To support this argument we remind that in a recent study\cite{hagelin2010}, indeed, a similar comparison of simulated versus measured wind speed obtained with a remote sensing instrument (a Generalized SCIDAR used for the wind speed measurements) provided a correlation better than 1 m$\cdot$s$^{-1}$ in the [0.5~km - 1~km] a.g.l. vertical range. 

\begin{table}
\begin{center}
\caption{\label{3_5_wind_speed} Maximum variability of the wind speed calculated at 4~km a.s.l inside a circle having a radius proportional to the wind speed observed at 4~km a.s.l. times 4 minutes.}
\vskip 0.2cm
\begin{tabular}{|c|c|c|c|}
\hline
Date & Hour (UT) & V$_{4km}$ (m$\cdot$s$^{-1}$)& $\Delta$V  (m$\cdot$s$^{-1}$)\\
\hline
1/8/2009 & 00:00 & 5  & 1.6 \\
11/11/2009 & 12:00 & 10 & 2 \\
19/11/2009 & 06:00& 5 & 1.4 \\
19/11/2009 & 12:00 & 10 & 1.6 \\
14/11/2009 & 12:00& 5 & 2 \\
\hline
\end{tabular}
\end{center}
\end{table}

\subsection{Wind speed and absolute temperature in the surface layer}
\label{sec:surface}
\begin{figure}
\centering
\includegraphics[width=14cm]{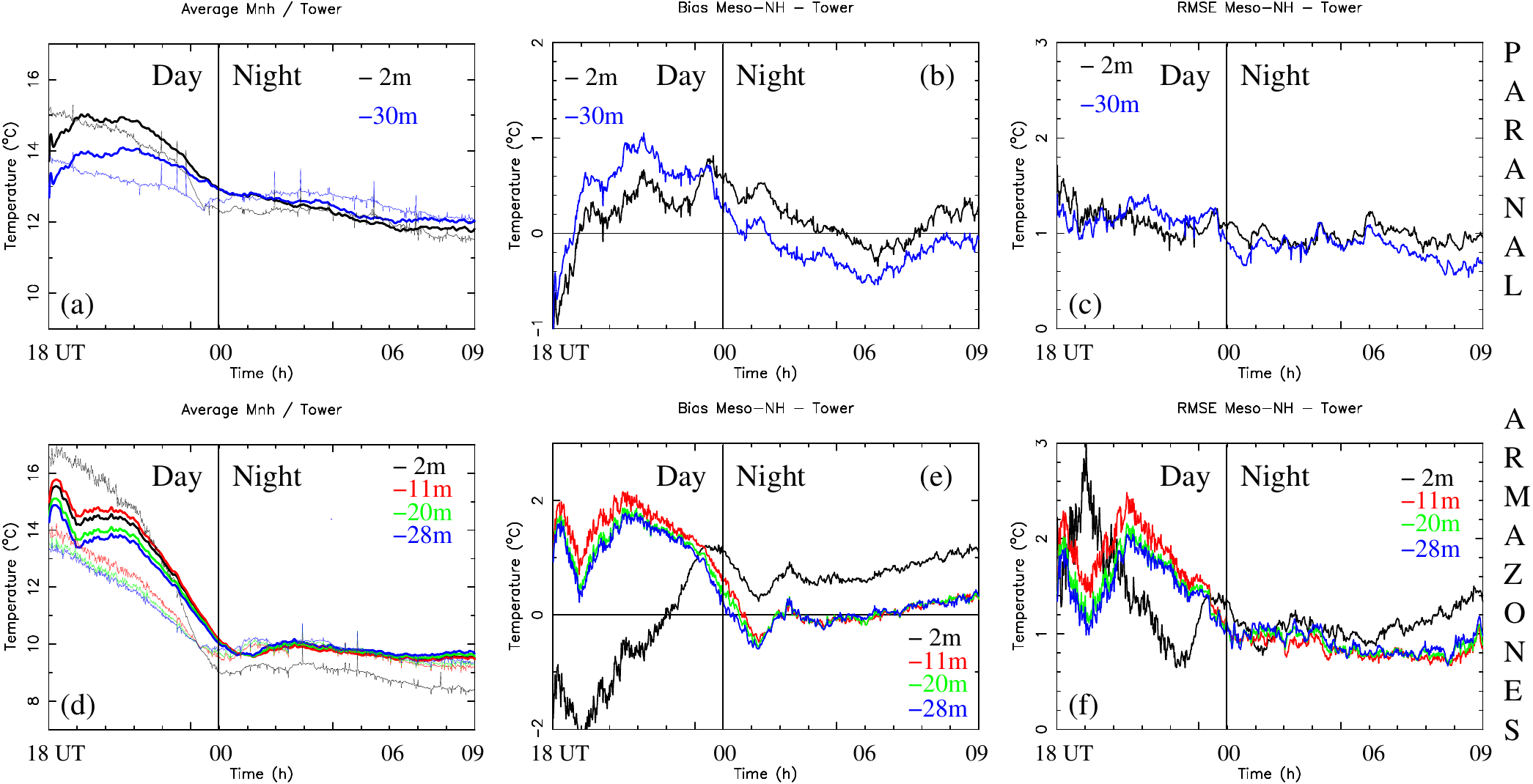}
\caption{\label{temp_surf_temp} Temporal evolution of the {\bf average} (left: (a),(d)), {\bf BIAS} (centre: (b),(e)) and {\bf RMSE} (right: (c),(f)) of the {\bf TEMPERATURE} observed and simulated on a statistical sample of {\bf 20 nights} and calculated above Cerro Paranal (top) and Cerro Armazones (bottom). In (a) and (d), simulations are in bold-line style, observations in thin-line style. Time x-axis is in UT time. The local nigh-time [20:00-05:00] LT is included in the [00:00-09:00] UT time.} 
\end{figure}

To quantify the model statistic reliability in reconstructing the meteorological surface parameters we estimate the bias and the RMSE of the nightly temporal evolution of the temperature and wind speed at each level for which observations are available at Cerro Paranal and Cerro Armazones. We perform the analysis on a statistic of {\bf 20 nights}. We selected all the nights of the Paranal site testing campaign of November/December 2007 on which observations (temperature and wind speed) near the surface are available on both sites plus a few others in which measurements above both sites were available. The bias and the RMSE are calculated for each instant with respect to N $\leq$ n (where n is the number of nights) which simply means that observed values are not always available at every minutes for every nights and in these cases the statistics is performed on a number of point inferior to the total number of nights. Fig.\ref{temp_surf_temp} shows the temporal evolution of the average, the bias and the RMSE (at different levels and above Cerro Paranal and Cerro Armazones) of the absolute temperature. Table \ref{par_temp} reports the corresponding statistical bias and the RMSE calculated with respect to the values observed all along the night. All simulations were initialized the day before at 18 UT and forced every 6 hours with the analyses from the ECMWF. Simulations finished at 09 UT of the simulated day (for a total duration of 15 hours). The statistics is computed only during night time, from 
00 UT to 09 UT. We neglect the first 6 hours related to the day time also because some spurious effect due to the model spin-up are present (in the first part of a simulation the model adapts itself to the orography). Looking at Fig.\ref{temp_surf_temp} and Table \ref{par_temp} we conclude that, above both Cerro Paranal and Cerro Armazones, we obtain excellent bias and RMSE values: the bias is well below 1$^{\circ}$C (at some heights well inferior to 0.2$^{\circ}$C) and, even more impressive, the RMSE is basically always inferior to 1$^{\circ}$C. The largest discrepancy is observed above Armazones at 2~m during the night (model overestimation of 0.76$^{\circ}$C). 
Looking at the temporal evolution of the temperature we can note that, for observations as well as for simulations and above the two sites, we can detect the inversion of the gradient of the temperature ($\frac{\partial T}{\partial h}$) near the surfaces when we pass from day to night time: $\frac{\partial T}{\partial h}$ $< $0 during the day-time, typical of convective regimes,  $\frac{\partial T}{\partial h}$ $>$ 0 during the night time, typical of stable regimes. Only above Armazones we note that only one level (11~m) presents a wrong gradient tendency with respect to the 2~meter level during both the day and night time period but the quantitative effect is minimum. The same model effect is observed also above Cerro Paranal. We conclude therefore that  the model, in general, well reconstructs the thermal structure near the surface even if there is still space for some improvements. As a further output of our analysis we note (Fig.\ref{temp_surf_temp} left side) that, on the sample of 20 night the temperature observed at 2~m at Cerro Armazones is typically almost 4$^{\circ}$C colder than at Cerro Paranal with a stronger gradient in the first 30 meters. This deserve a deeper investigation because, in case this should be confirmed on a climatological scale, this should it should be an indication that at Cerro Armazones the turbulence surface layer is typically thinner than at Cerro Paranal with a stronger strength. This correlation between the thermal gradient of the surface layer, the thickness of the surface layer and the optical turbulence inside the surface layer has been put in evidence in one of our previous study (Lascaux et al. 2011\cite{Lascaux2011}) above the internal Antarctic plateau.

\begin{table}
\begin{center}
\caption{\label{par_temp} {\bf BIAS} (simulations minus observations) and {\bf RMSE} estimations of the {\bf TEMPERATURE} between
Meso-NH simulations and observations, at Cerro Paranal (left) and Cerro Armazones (right). 
Units in $^o$C. The statistical sample is made by each individual couple of values for each instant for which we can compare simulations versus observations for each night (20).}
\vskip 0.2cm
\begin{tabular}{|c|c|c|c||c|c|c|c|c|c|}
\hline
\multicolumn{2}{|c|}{PARANAL} & 2~m  & 30~m & \multicolumn{2}{|c|}{ARMAZONES} & 2~m  & 11~m   & 20~m & 28~m  \\
\hline
\multirow{2}{*}{3dom} & BIAS  & 0.13 & -0.19  & \multirow{2}{*}{3dom} & BIAS    & 0.76   & 0.04   & 0.03 & 0.01  \\
                                    & RMSE  & 0.97   & 0.84    &                       & RMSE    & 1.10   & 0.85     & 0.89   & 0.92    \\  
\hline
\end{tabular}
\end{center}
\end{table}

\begin{figure}
\centering
\includegraphics[width=14cm]{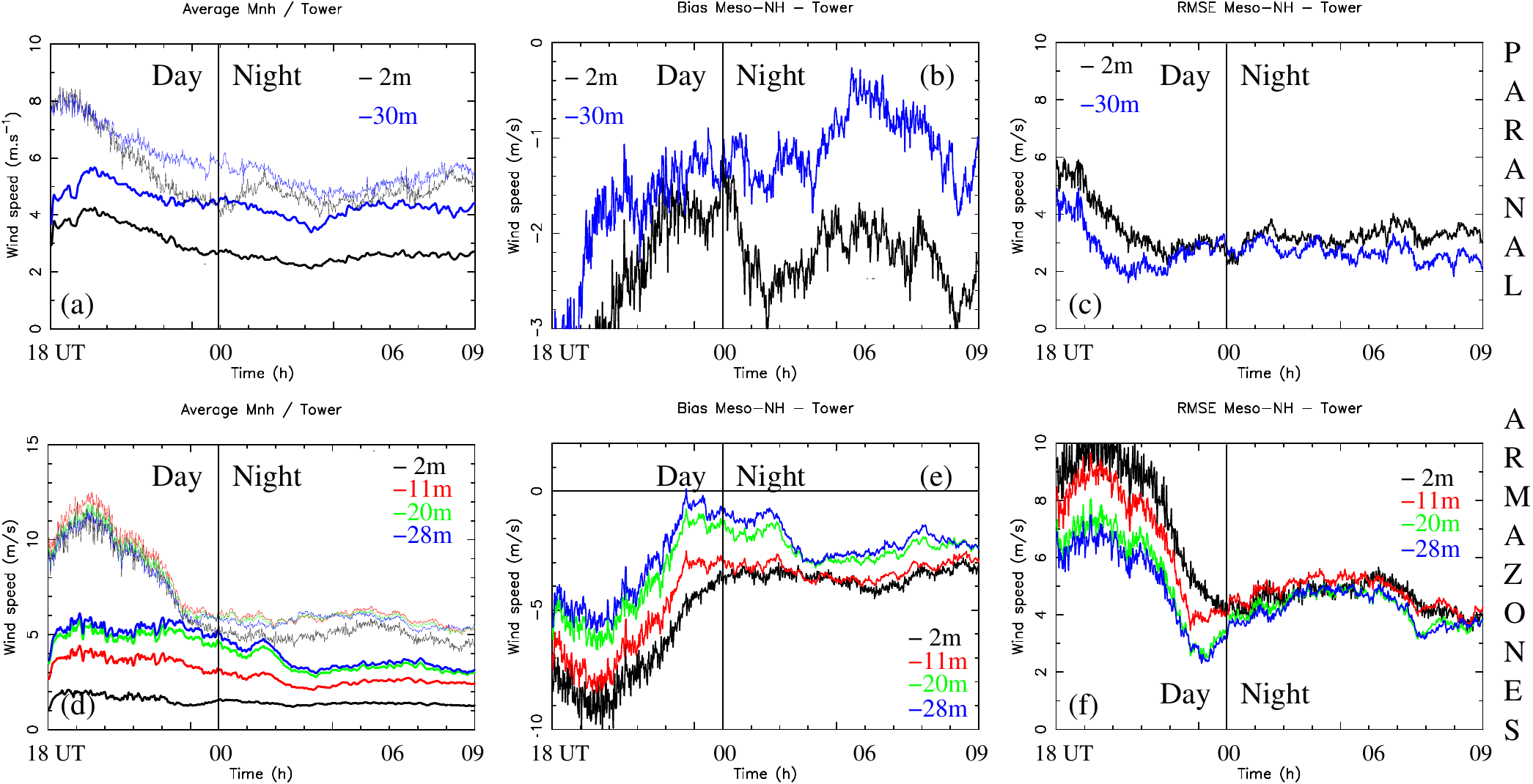}
\caption{\label{temp_surf_wind} Temporal evolution of the {\bf average} (left: (a),(d)), {\bf BIAS} (centre: (b),(e)) and {\bf RMSE} (right: (c),(f)) of the {\bf WIND SPEED} observed and simulated on a statistical sample of {\bf 20 nights} and calculated above Cerro Paranal (top) and Cerro Armazones (bottom). In (a) and (d), simulations are in bold-line style, observations in thin-line style. Time x-axis is in UT time. The local nigh-time [20:00-05:00] LT is included in the [00:00-09:00] UT time.} 
\end{figure}

\begin{table}
\begin{center}
\caption{\label{par_wind} {\bf BIAS} (simulations minus observations) and {\bf RMSE} estimations of the {\bf WIND SPEED} between
Meso-NH simulations and observations, at Cerro Paranal (left) and Cerro Armazones (right). 
Units in m$\cdot$s$^{-1}$. The statistical sample is made by each individual couple of values for each instant for which we can compare simulations versus observations for each night (20).}
\vskip 0.2cm
\begin{tabular}{|c|c|c|c||c|c|c|c|c|c|}
\hline
\multicolumn{2}{|c|}{PARANAL} & 2~m  & 30~m & \multicolumn{2}{|c|}{ARMAZONES} & 2~m  & 11~m   & 20~m & 28~m  \\
\hline
\multirow{2}{*}{3dom} & BIAS  & -2.23 & -1.07  & \multirow{2}{*}{3dom} & BIAS    &  -3.59  & -3.33 &-2.38  & -2.06  \\
                                    & RMSE  &  3.24  & 2.68   &                       & RMSE    & 4.61   &  4.81  &  4.29  &  4.25 \\  
\hline
\end{tabular}
\end{center}
\end{table}

\begin{figure}
\centering
\includegraphics[width=13cm]{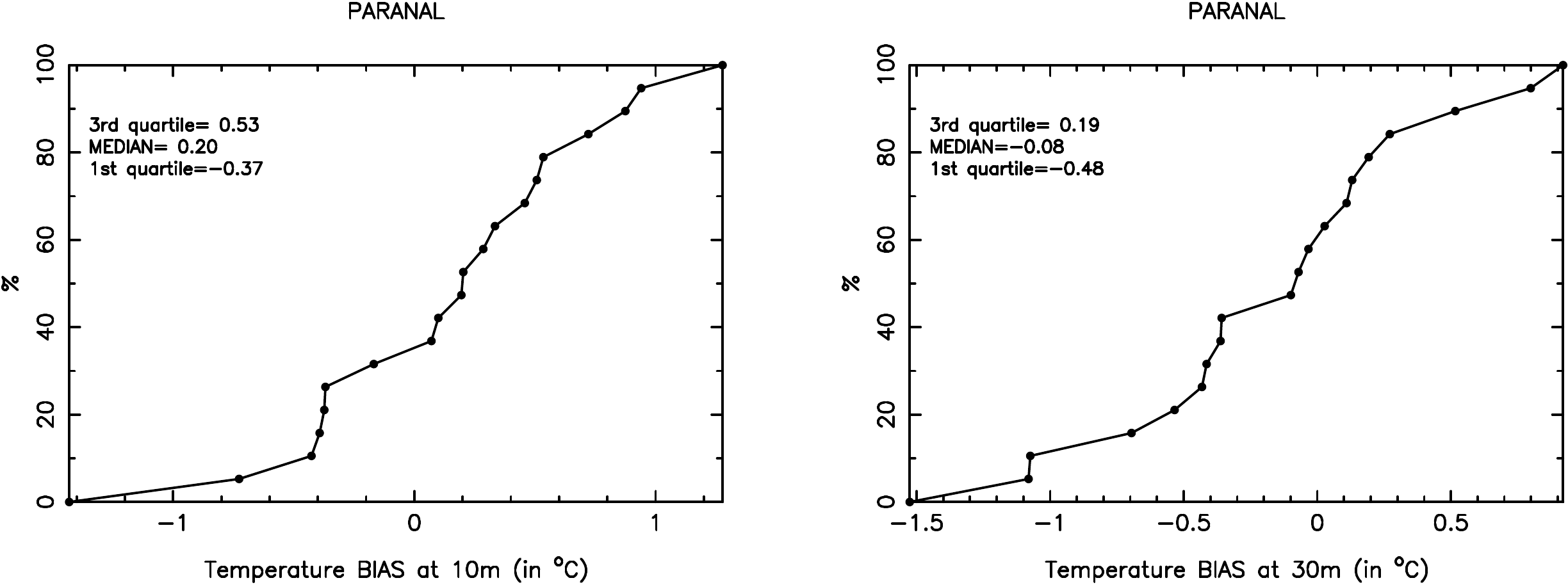}
\includegraphics[width=13cm]{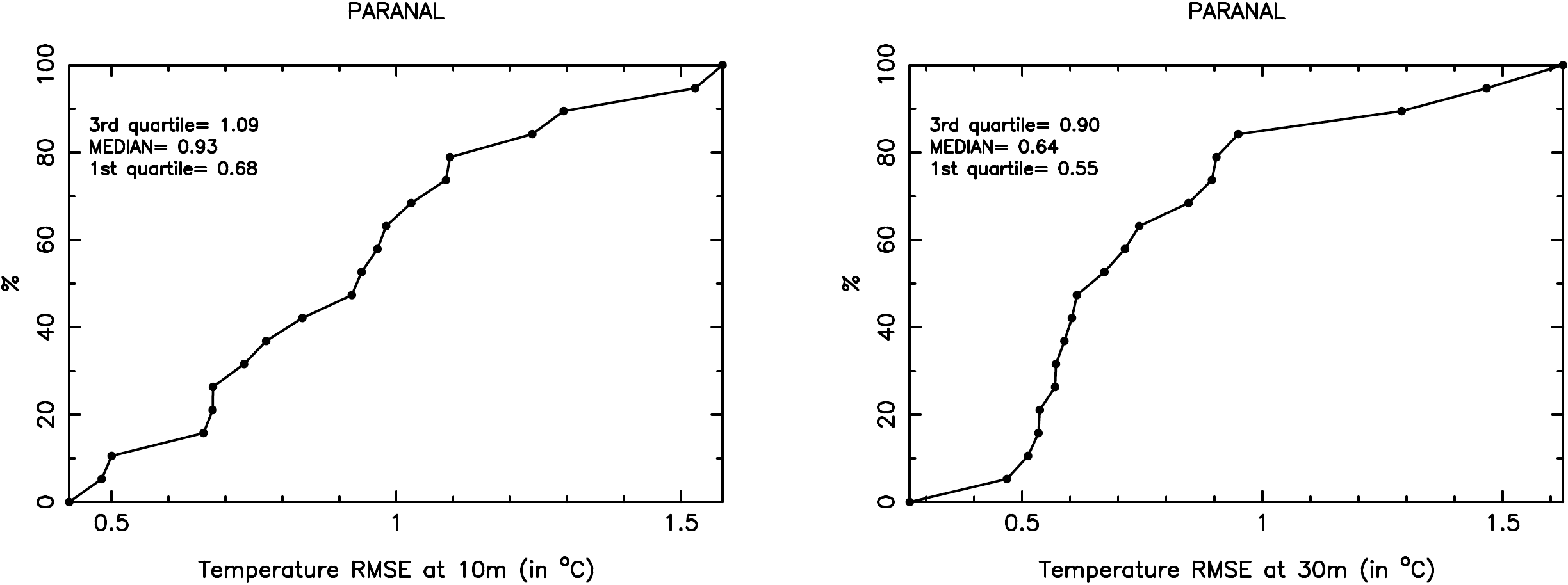}
\caption{\label{par_bias_rmse_ind_nights} Cumulative distribution of the temperature at 10~m and 30~m for the BIAS and the RMSE at Cerro Paranal. The BIAS and the RMSE are calculated with respect to the values related to the individual nights i.e. the statistical sample is 20 nights.} 
\end{figure}

\begin{figure}
\centering
\includegraphics[width=13cm]{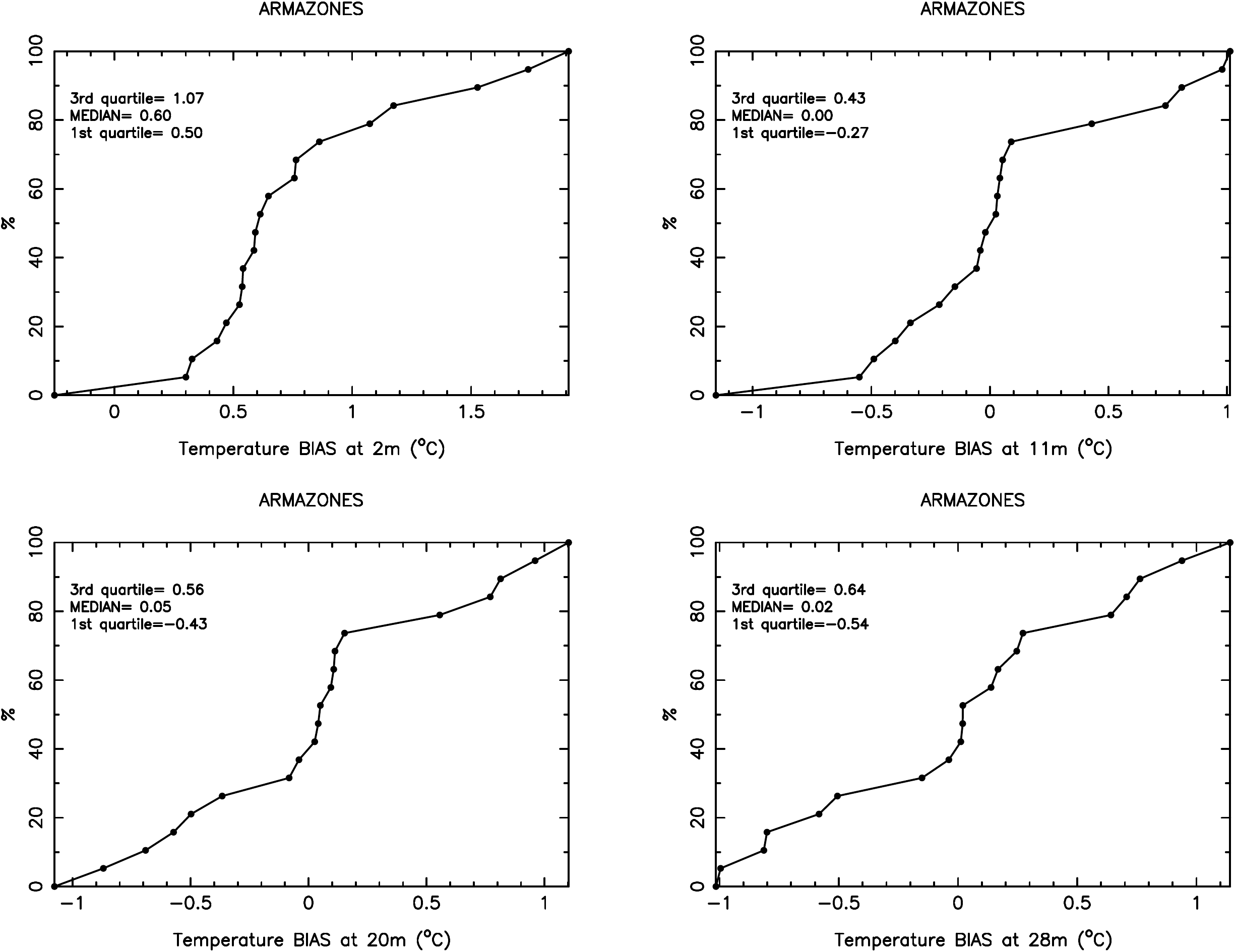}
\caption{\label{arm_bias_ind_nights} Cumulative distribution of the temperature at 2~m, 11~m, 20~m and 28~m for the BIAS at Cerro Armazones. The BIAS is calculated with respect to the values related to the individual nights i.e. the statistical sample is 20 nights.} 
\end{figure}
\begin{figure}
\centering
\includegraphics[width=13cm]{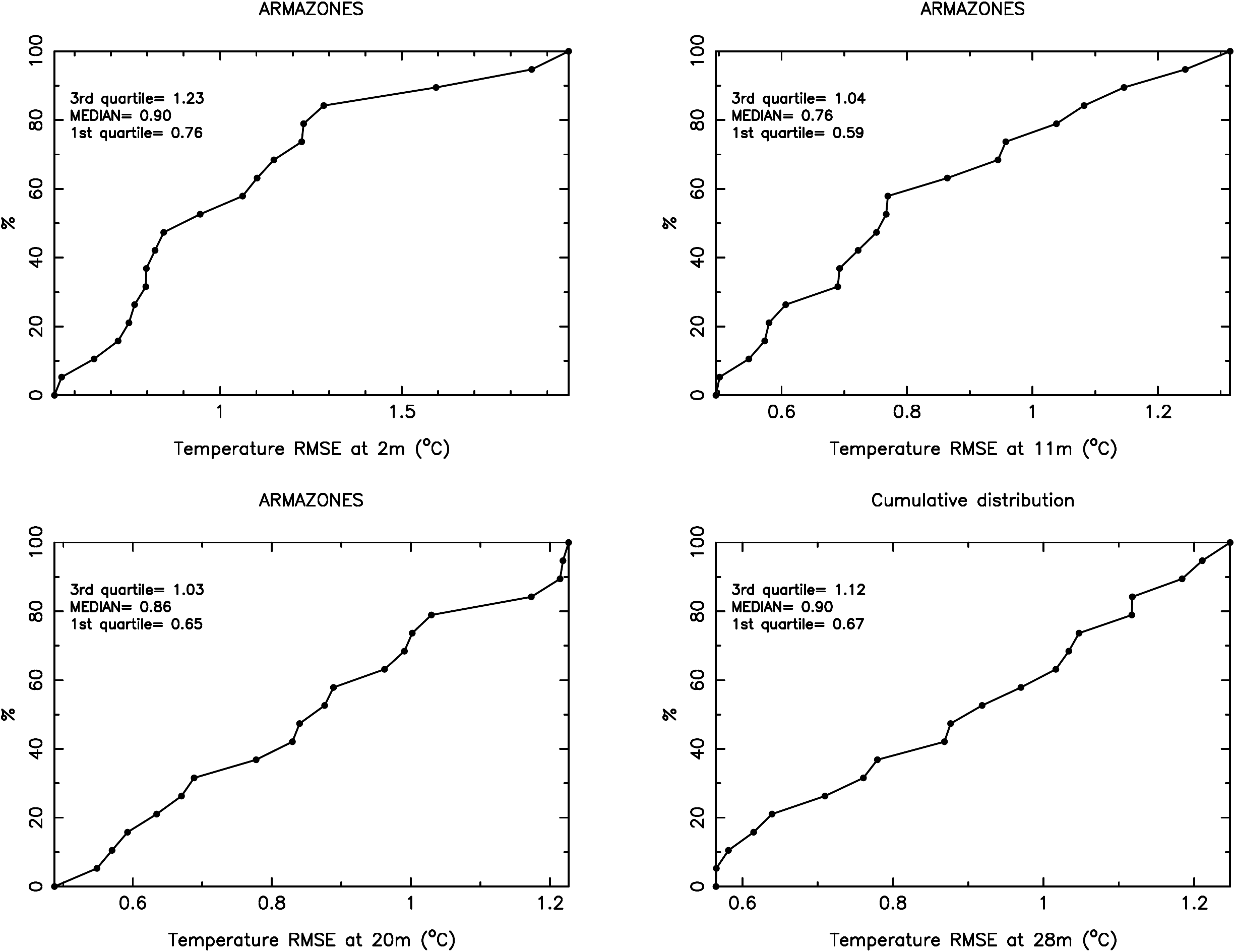}
\caption{\label{arm_rmse_ind_nights_rmse} Cumulative distribution of the temperature at 2~m, 11~m, 20~m and 28~m for the RMSE at Cerro Armazones. The RMSE is calculated with respect to the values related to the individual nights i.e. the statistical sample is 20 nights.} 
\end{figure}
Fig.\ref{temp_surf_wind} shows the temporal evolution of the average, the bias and the RMSE (at different levels and above Cerro Paranal and Cerro Armazones) of the wind speed. Table \ref{par_wind} reports the corresponding statistical bias and the RMSE calculated with respect to the values observed all along the night. At Cerro Paranal the maximum bias is 2.23 m$\cdot$s$^{-1}$ while at Armazones is 3.59 m$\cdot$s$^{-1}$. The RMSE has maximum values respectively of 3.24 and 4.81 m$\cdot$s$^{-1}$ above Cerro Paranal and Cerro Armazones. We note a general tendency of the model in underestimating the wind speed all along night, particularly at the first level (2~m) even if the model seems to show a better behavior above Cerro Paranal than Cerro Armazones. Even if the bias and the RMSE are still small in absolute terms, considering the typical absolute values of the wind speed at this heights, we can have a not negligible relative error. In a further study Lascaux \& Masciadri\cite{lascaux2012} presented in this Conference, we performed an analog statistical analysis with a different model configuration (five imbricated models with the higher horizontal resolution of 100~m) to check if the model performances in reconstructing the wind speed near the surface in proximity of these summits can be improved. We found that such a new configuration definitely and substantially improves the model performances in reconstructing the wind speed near the surface of around the 50$\%$. 

Which are the model performances in reconstructing these parameters night by night ? Fig.\ref{par_bias_rmse_ind_nights} shows the cumulative distribution of the BIAS and the RMSE of simulations versus observations for the temperature at 2~m and 30~m achieved by the model at Cerro Paranal night by night. Fig.\ref{arm_bias_ind_nights} and Fig.\ref{arm_rmse_ind_nights_rmse} show respectively the bias and the RMSE obtained at Cerro Armazones at each height: 2~m, 11~m, 20~m and 28~m. We note that the level of the model accuracy is very satisfactory above both sites. The median value of the bias is always within 0.60$^{\circ}$C with a maximum absolute value of the quartiles equal to 1.07$^{\circ}$C. The median value of the RMSE is within 0.93$^{\circ}$C if we take into account all the model levels analyzed with a maximum third quartile of 1.23$^{\circ}$C.


\section{CONCLUSIONS}
\label{sec:concl} 

In this contribution we present the preliminary results obtained in the context of an extended feasibility study (MOSE project) aiming at performing a feasibility study for the forecast of the optical turbulence and the atmospherical parameters above Cerro Paranal and Cerro Armazones. We focus here on the atmospheric parameters in the free atmosphere as well as at the surface. We proved that the Meso-Nh mesoscale model, using a configuration made by three imbricated domains (horizontal resolution of 10~km, 2.5~km and 0.5~km), provides vertical distribution (from the ground up to $\sim$ 20km) of the main classical atmospheric parameters (wind speed and direction, temperature and relative humidity) with excellent levels of correlation with observations. The residual discrepancies are mainly due to the initial conditions. We showed that, at present, the Meso-Nh mesoscale model can be a very useful tool to calculate and predict the temporal evolution of the wind speed for the calculation of the wavefront coherence time ($\tauO$). This method offers a very advantageous solution in terms of accuracy and temporal coverage. 

Near the ground, in proximity of the surface, the forecast of the temperature calculated at different heights in the first 30 meters shows excellent performances above both sites (Paranal and Armazones) with a statistical bias with respect to observations whose median value is well smaller than 1$^{\circ}$C (at some heights well inferior to 0.2$^{\circ}$C) and a RMSE always inferior to 1$^{\circ}$C. The statistical sample is very rich considering a set of couples of values sampled at a frequency of 1 minutes,  extended on 9 hours (night time) for a total number of 20 nights. A similar analysis performed for the wind speed indicated that the model shows, in this case, a tendency in underestimating the wind speed, particularly at the first level. This effect is more evident for a strong wind speed. The biases are not important in absolute terms: order of 2.23 and 3.59 m$\cdot$s$^{-1}$ at 2~m and a little weaker for the higher levels at 20~m, 28~m and 30~m (order of 2-2.5 m$\cdot$s$^{-1}$). However the model tendency is systematic and it is therefore reasonable to investigate on how to improve the model behavior in this context. In a further dedicated study (Lascaux \& Masciadri\cite{lascaux2012}) we presented in this conference how to overcome this limitation.

Beside a statistical analysis performed on a rich statistical samples, we presented in this contribution a detailed analysis of the model performances in forecasting the atmospherical parameters night by night. For what concerns the vertical distribution, the level of accuracy shown by the model is impressive. In basically all the cases studied the model is able to reconstruct the different features of the wind speed observed in three different instants during the nights (example in Fig.\ref{radio_mnh}). For what concerns the state of the atmosphere in proximity of the surface we proved that the median values of the biases (calculated for each single night and calculated at different heights: 2~m, 11~m, 20~m, 28~m, 30~m) is within 0.60$^{\circ}$C above Cerro Paranal and Cerro Armazones with a maximum absolute value of the quartiles of 1.07$^{\circ}$C. The median value of the RMSEs is within 0.93$^{\circ}$C if we take into account all the model levels analyzed with a maximum third quartile of 1.23$^{\circ}$C. We conclude, therefore, that the Meso-Nh model appears as an extremely useful system to predict the temperature near the ground with excellent levels of accuracy. Such a system can play a fundamental role in monitoring and forecasting the thermal stratification of the first few tens of meters above the ground and to infer the status of thermal stability i.e. the value of the $\CN2$ in the shallow surface layer. This definitely can play a fundamental role in the service mode operation of scientific programs.

\acknowledgments     
 
Meteorological data-set from the Automatic Weather Station (AWS) and mast at Cerro Armazones are from the Thirty Meter Telescope Site Testing - Public Database Server\cite{Schoeck2009}.
Meteorological data-set from the AWS and mast at Cerro Paranal are from ESO Astronomical Site Monitor (ASM - Doc.N. VLT-MAN-ESO-17440-1773). We are very grateful to the whole staff of the TMT Site Testing Working Group for providing information about their data-set as well as to Marc Sarazin for his constant support to this study and for providing us the ESO data-set used in this study and Florian Kerber for his valuable support in accessing at radio-soundings data-set. Simulations are run partially on the HPCF cluster of the European Centre for Medium Weather Forecasts (ECMWF) - Project SPITFOT. This study is co-funded by the ESO contract: E-SOW-ESO-245-0933.



\end{document}